\documentstyle[12pt]{article}
\newcommand{\ehat}{ \hat U_{\epsilon} }

\newcommand{\define}{ \stackrel{\triangle}{=} }

\def\be{\begin{equation}}
\def\ee{\end{equation}}
\def\ba{\begin{array}}
\def\ea{\end{array}}

\def\d4{{\rm d}^4}

\begin{document}
\title{\bf Unification of Non-Abelian $SU(N)$ Gauge Theory and
                Gravitational Gauge Theory }
\author{{Ning Wu}
\thanks{email address: wuning@mail.ihep.ac.cn}
\\
\\
{\small Institute of High Energy Physics, P.O.Box 918-1,
Beijing 100039, P.R.China}
\thanks{mailing address}}
\maketitle
\vskip 0.8in

~~

\vskip 0.8in

\begin{abstract}
In this paper, a general theory on unification of non-Abelian
$SU(N)$ gauge interactions and gravitational interactions
is discussed. $SU(N)$ gauge interactions and gravitational 
interactions are formulated on the similar basis and are 
unified in a semi-direct product group $GSU(N)$. Based on
this model, we can discuss unification of fundamental 
interactions of Nature. 
\\

\end{abstract}

\newpage

\Roman{section}

\section{Introduction}

It is known that there are four kinds of fundamental interactions
in Nature, which are strong interactions, electromagnetic
interactions, weak interactions and gravitational interactions.
All these fundamental interactions can be described by gauge field
theories, which can be regarded as the common nature of all these
fundamental interactions. It provides us the possibility to 
unify different kinds of
fundamental interactions in the framework of gauge theory. The first
unification of fundamental interactions in human history is the
unification of electric interactions and magnetic interactions,
which is made by Maxwell in 1864. Now, we know that electromagnetic
theory is a $U(1)$ Abelian gauge theory. In 1921, H.Weyl tried to
unify electromagnetic interactions and gravitational interactions
in a theory which has local scale invariance\cite{1,2}. Weyl's
original work is not successful, however in his great attempt, he
introduced one of the most important concept in modern physics:
gauge transformation and gauge symmetry. After the foundation
of quantum mechanics, V.Fock, H.Weyl and W.Pauli found that quantum
electrodynamics is a $U(1)$ gauge invariant theory\cite{3,4,5}.
\\

In 1954, Yang and Mills proposed non-Abelian gauge field theory\cite{6}.
Soon after, non-Abelian gauge field theory is applied to elementary
particle theory. In about 1967 and 1968, using the spontaneously
symmetry breaking and Higgs mechanism\cite{7,8,9,10,11,12,13,14},
S.Weinberg and A.Salam proposed the unified electroweak theory,
which is a $SU(2) \times U(1)$ gauge theory\cite{15,16,17}. 
The predictions of unified electroweak theory have been confirmed
in a large number of experiments, and the intermediate gauge
bosons $W^{\pm}$ and $Z^0$ which are predicted by unified
electroweak model are also found in experiments. From nineteen
seventies, physicist begin studying Grand Unified theories which try
to unify strong, electromagnetic and weak interactions in a simple
group. At that time, $SU(5)$ model\cite{18,19},
$SO(10)$ model\cite{20,21,22}, $E_6$ model\cite{23,24,25}
and other models\cite{26,27,28} are proposed. In these attempts,
gravitational interactions are not considerdd.\\

Gauge treatment of gravity was suggested immediately after the
gauge theory birth itself\cite{29,30,31}. In the traditional gauge
treatment of gravity, Lorentz group is localized, and the
gravitational field is not represented by gauge
potential\cite{32,33,34}. It is represented by metric field. 
The theory has beautiful mathematical forms, but up to now, its
renormalizability is not proved. In other words,
it is conventionally considered to be 
perturbatively non-renormalizable.
Recently, some new attempts were proposed to use Yang-Mills theory
to reformulate quantum gravity\cite{35,36,37,38,38a}. In these new
approaches, the importance of gauge fields is emphasized. Some
physicists also try to use gauge potential to represent
gravitational field, some suggest that we should pay more
attention on translation group. Recently, Wu proposed a new quantum
gauge theory of gravity which is a renormalizable quantum
gravity\cite{39}. Based on gauge principle, space-time translation
group is selected to be the gravitational gauge group. After
localization of gravitational gauge group, the gravitational field
appears as the corresponding gauge potential. In this paper,
we will discuss unification of fundamental interactions which is
based on gravitational gauge theory. In literature \cite{41},
Wu proposed a model in which $U(1)$ gauge theory is consistently
unified with gravitational gauge theory
in a semi-direct product group. In this paper, we
will generalize the unification from $U(1)$ gauge group to
$SU(N)$ gauge group, i.e., the unification of general non-Abelian
$SU(N)$ gauge interactions and gravitational gauge interactions
is studied. If $N=3$, this theory is just
the unified theory of strong and gravitational interactions.
\\

\section{Lagrangian}
\setcounter{equation}{0}    

The generators of $SU(N)$ group is denoted as $T_a$, they
satisfies
\be \label{3.1}
\lbrack T_a~~,~~ T_b \rbrack = i f_{abc} T_c,
\ee
\be \label{3.2}
Tr( T_a  T_b )  = K \delta_{ab}.
\ee
The $SU(N)$ non-Abelian gauge field is denoted as $A_{\mu}$, which is
an element of $SU(N)$ Lie algebra,
\be \label{3.3}
A_{\mu}(x) = A_{\mu}^{a}(x) T_a,
\ee
where $A_{\mu}^{a}(x)$ are component fields. \\

Because an arbitrary element $U(x)$ of $SU(N)$ group does not commute
with an arbitrary element $\ehat$ of gravitational gauge group,
\be \label{3.4}
\lbrack U(x)~~,~~ \ehat \rbrack \not= 0.
\ee
The product group of $SU(N)$ group and gravitational group is not
direct product group, but semi-direct product group, which we will
denoted as $GSU(N)$
\be \label{3.5}
GSU(N) \define
SU(N) \otimes_s Gravitational~Gauge~Group.
\ee
An arbitrary element of $GSU(N)$ is denoted as $g(x)$, which is
defined by
\be \label{3.6}
g(x) \define \ehat \cdot U(x).
\ee
The gauge covariant derivative of $GSU(N)$ group is
\be \label{3.7}
{\mathbf D}_{\mu} \define \partial_{\mu}
- i g C_{\mu} - i g_s A_{\mu} = D_{\mu} -ig_s A_{\mu},
\ee
where $g_s$ is the coupling constant of non-Abelian $SU(N)$ gauge
interactions, $C_{\mu}$ is the gravitational gauge field 
and $D_{\mu}$ is the gravitational gauge 
covariant derivative
\be \label{3.701}
D_{\mu} = \partial_{\mu} - i g C_{\mu} (x).
\ee  
Gravitational gauge field which is a vector in gauge
group space,
\be \label{3.702}
C_{\mu}(x) = C_{\mu}^{\alpha} (x) \hat{P}_{\alpha},
\ee
where $\hat{P}_{\alpha} = - i \frac{\partial}{\partial x^{\alpha}}$ 
is the generator of gravitational gauge group.
\\

The field strength of non-Abelian gauge field $A_{\mu}$ is
\be \label{3.8}
A_{\mu\nu} = (D_{\mu} A_{\nu}) - (D_{\nu} A_{\mu})
-ig_s \lbrack A_{\mu} ~~,~~ A_{\nu} \rbrack.
\ee
$A_{\mu\nu}$ is also an element of $SU(N)$ Lie algebra,
\be \label{3.9}
A_{\mu\nu}(x) = A_{\mu\nu}^{a}(x) T_a,
\ee
where
\be \label{3.10}
A^a_{\mu\nu} = (D_{\mu} A^a_{\nu}) - (D_{\nu} A^a_{\mu})
+ g_s f_{abc} A^b_{\mu}  A^c_{\nu}.
\ee
$A_{\mu\nu}$ is not a $SU(N)$ gauge covariant field strength. 
In order to define $SU(N)$ gauge covariant field strength, we 
need a matrix $G$ which is given by
\be \label{2.11}
G = (G_{\mu}^{\alpha}) = ( \delta_{\mu}^{\alpha} - g C_{\mu}^{\alpha} )
= I - gC.
\ee
Its inverse matrix is denoted as $G^{-1}$,
\be \label{2.12}
G^{-1} = \frac{1}{I - gC} = (G^{-1 \mu}_{\alpha}).
\ee
They satisfy the following relations,
\be \label{2.13}
G_{\mu}^{\alpha} G^{-1 \nu}_{\alpha} = \delta_{\mu}^{\nu},
\ee
\be \label{2.14}
G_{\beta}^{-1 \mu} G^{ \alpha}_{\mu} = \delta_{\beta}^{\alpha}.
\ee
It can be proved that
\be \label{2.15}
D_{\mu}= G_{\mu}^{\alpha} \partial_{\alpha}.
\ee
The  field strength of gravitational gauge field is defined by
\be \label{2.16}
F_{\mu\nu} \define \frac{1}{-ig} \lbrack D_{\mu}~~,~~D_{\nu} \rbrack.
\ee
Its explicit expression is
\be \label{2.17}
F_{\mu\nu}(x) = \partial_{\mu} C_{\nu} (x)
-\partial_{\nu} C_{\mu} (x)
- i g C_{\mu} (x) C_{\nu}(x)
+ i g C_{\nu} (x) C_{\mu}(x).
\ee
$F_{\mu\nu}$ is also a vector in gauge group space,
\be \label{2.18}
F_{\mu\nu} (x) = F_{\mu\nu}^{\alpha}(x) \cdot \hat{P}_{\alpha},
\ee
where
\be \label{2.19}
F_{\mu\nu}^{\alpha} = \partial_{\mu} C_{\nu}^{\alpha}
-\partial_{\nu} C_{\mu}^{\alpha}
-  g C_{\mu}^{\beta} ( \partial_{\beta} C_{\nu}^{\alpha})
+  g C_{\nu}^{\beta} ( \partial_{\beta} C_{\mu}^{\alpha}).
\ee
\\

$SU(N)$ Gauge covariant field strength is defined by
\be \label{3.11}
{\mathbf A}_{\mu\nu} = A_{\mu\nu} + g G^{-1 \lambda}_{\sigma}
A_{\lambda} F_{\mu\nu}^{\sigma} =
{\mathbf A}^a_{\mu\nu} T_a,
\ee
where
\be \label{3.12}
{\mathbf A}^a_{\mu\nu} = A^a_{\mu\nu} + g G^{-1 \lambda}_{\sigma}
A^a_{\lambda} F_{\mu\nu}^{\sigma} .
\ee
The lagrangian density ${\cal L}_0$ is given by,
\be \label{3.13}
\begin{array}{rcl}
{\cal L}_0 &=&
-  \bar{\psi}
\lbrack \gamma^{\mu} (D_{\mu} -i g_s A_{\mu} ) + m
\rbrack \psi
-\frac{1}{4} \eta^{\mu \rho} \eta^{\nu \sigma}
{\mathbf A}^a_{\mu \nu } {\mathbf A}^a_{ \rho \sigma } \\
&&\\
&&- \frac{1}{4} \eta^{\mu \rho} \eta^{\nu \sigma}
g_{\alpha \beta }
F_{\mu \nu}^{\alpha} F_{\rho \sigma}^{\beta},
\end{array}
\ee
where
\be \label{2.27}
g_{\alpha \beta} \define \eta_{\mu \nu}
(G^{-1})_{\alpha}^{\mu} (G^{-1})_{\beta}^{\nu}.
\ee
\\

The full lagrangian density ${\cal L}$ is defined by
\be \label{3.14}
{\cal L} = J(C) {\cal L}_0,
\ee
where
\be \label{2.26}
J(C) = \sqrt{- {\rm det} g_{\alpha \beta} }
\ee
is a special factor to resume gravitational gauge 
symmetry of the system. 
The action is
\be \label{3.15}
S  = \int \d4 x {\cal L} = \int\d4 x J(C) {\cal L}_0.
\ee
\\

\section{Gauge Symmetry}
\setcounter{equation}{0}

Now, we discus symmetry of the system. Under $SU(N)$ gauge
transformations, gravitational gauge field $C_{\mu}(x)$
is kept unchanged. Therefore, $F_{\mu\nu}^{\alpha}$,
$G^{\alpha}_{\mu}$, $G^{-1 \mu}_{\alpha}$ , $D_{\mu}$
and $J(C)$
are not changed under local $SU(N)$ gauge transformations.
Other fields and operators transform as
\be \label{3.16}
\psi  \to \psi' =  (U(x) \psi),
\ee
\be \label{3.17}
A_{\mu} \to A'_{\mu} = U(x) A_{\mu} U^{-1}(x)
- \frac{1}{ig_s} U(x) (D_{\mu} U^{-1}(x)) ,
\ee
\be \label{3.18}
A_{\mu\nu} \to A'_{\mu\nu} = U(x) A_{\mu\nu} U^{-1}(x)
+ \frac{g}{ig_s} F_{\mu\nu}^{\sigma}
U(x) (\partial_{\sigma} U^{-1}(x)) ,
\ee
\be \label{3.19}
{\mathbf A}_{\mu\nu} \to {\mathbf A}'_{\mu\nu}
= U(x) {\mathbf A}_{\mu\nu} U^{-1}(x).
\ee
Using all these relations, we can prove that the lagrangian
density ${\cal L}_0$ does not change under local $SU(N)$
gauge transformations
\be \label{3.20}
{\cal L}_0 \to {\cal L}'_0 = {\cal L}_0.
\ee
Because both integration measure $\d4 x$ and $J(C)$ are not
changed under non-Abelian $SU(N)$ gauge transformation, the action
is invariant under $SU(N)$ gauge transformation. Therefore, 
the system has local $SU(N)$ gauge symmetry.\\

Under local gravitational gauge transformation, the
transformations of various fields and operators are
\be \label{3.21}
\psi \to \psi ' = (\ehat \psi),
\ee
\be \label{3.22}
\bar\psi \to \bar\psi ' = (\ehat \bar\psi),
\ee
\be \label{3.23}
A_{\mu} \to A'_{\mu} = \ehat A_{\mu} \ehat^{-1},
\ee
\be \label{3.24}
C_{\mu} \to C'_{\mu} = \ehat C_{\mu} \ehat^{-1}
- \frac{1}{ig} \ehat (\partial_{\mu}  \ehat^{-1}),
\ee
\be \label{3.241}
g_{\alpha\beta} \to g'_{\alpha\beta} 
= \Lambda_{\alpha}^{~\alpha_1} \Lambda_{\beta}^{~\beta_1}
(\ehat g_{\alpha_1 \beta_1} ),
\ee
\be \label{3.25}
{\mathbf D}_{\mu} \to {\mathbf D}'_{\mu}
= \ehat {\mathbf D}_{\mu} \ehat^{-1},
\ee
\be \label{3.26}
A_{\mu\nu} \to A'_{\mu\nu} = \ehat A_{\mu\nu} \ehat^{-1},
\ee
\be \label{3.27}
F_{\mu\nu}^{\sigma} \to F_{\mu\nu}^{\prime\sigma}
= \Lambda^{\sigma}_{~\rho} \ehat F_{\mu\nu}^{\rho} \ehat^{-1},
\ee
\be \label{3.28}
G^{\alpha}_{\mu} \to G^{\prime\alpha}_{\mu}
= \Lambda^{\alpha}_{~\beta} \ehat G^{\beta}_{\mu} \ehat^{-1},
\ee
\be \label{3.29}
G_{\alpha}^{-1\mu} \to G_{\alpha}^{\prime -1 \mu}
= \Lambda_{\alpha}^{~\beta} \ehat G_{\beta}^{-1\mu} \ehat^{-1},
\ee
\be \label{3.30}
{\mathbf A}_{\mu\nu} \to {\mathbf A}'_{\mu\nu}
= \ehat {\mathbf A}_{\mu\nu} \ehat^{-1},
\ee
\be \label{3.31}
J(C) \to J'(C') = J \cdot \ehat J(C) \ehat^{-1},
\ee
where $J$ is the Jacobian of the corresponding transformation
which is given by
\be \label{3.3101}
J = det \left (\frac{\partial (x - \epsilon)^{\mu}}
{\partial x^{\nu}}\right),
\ee
and $\Lambda^{\alpha}_{~\beta}$ and 
$\Lambda_{\alpha}^{~\beta}$ are the transformation
matrices which are given by \cite{39,40}:
\be
\Lambda^{\alpha}_{~~\beta} =
\frac{\partial x^{\alpha}}{\partial ( x - \epsilon (x) )^{\beta}},
\label{3.3102}
\ee
\be
\Lambda_{\alpha}^{~~\beta} =
\frac{\partial ( x - \epsilon (x))^{\beta}}{\partial x^{\alpha}}.
\label{3.3103}
\ee
Using all these relations and the following relation
\be \label{3.3102}
\int {\rm d}^4 x J(\ehat f(x)) = \int {\rm d}^4 x f(x),
\ee
where $f(x)$ is an arbitrary function, we can prove that
\be \label{3.32}
{\cal L}_0 \to {\cal L}'_0 =( \ehat {\cal L}_0 ),
\ee
\be \label{3.33}
{\cal L} \to {\cal L}' = J( \ehat {\cal L} ),
\ee
\be \label{3.34}
S \to S' = S.
\ee
Therefore, the system has local gravitational gauge symmetry.\\

Combining above results on local $SU(N)$ gauge transformations and local
gravitational gauge transformations, we know that under general
$GSU(N)$ gauge transformation $g(x)$, transformations of various
fields and operators are
\be \label{3.35}
\psi \to \psi ' = (g(x) \psi),
\ee
\be \label{3.36}
\bar\psi \to \bar\psi ' = (\ehat (\bar\psi U^{\dag}(x) )),
\ee
\be \label{3.37}
A_{\mu} \to A'_{\mu} = g(x) \left\lbrack A_{\mu}
- \frac{1}{ig_s} (D_{\mu} U^{-1}(x)) U(x)
\right\rbrack g^{-1}(x),
\ee
\be \label{3.38}
C_{\mu} \to C'_{\mu} = \ehat C_{\mu} \ehat^{-1}
- \frac{1}{ig} \ehat (\partial_{\mu}  \ehat^{-1}),
\ee
\be \label{3.39}
{\mathbf D}_{\mu} \to {\mathbf D}'_{\mu}
= g(x) {\mathbf D}_{\mu} g^{-1}(x),
\ee
\be \label{3.40}
A_{\mu\nu} \to A'_{\mu\nu} = g(x) \left\lbrack  A_{\mu\nu}
+\frac{g}{i g_s} F_{\mu\nu}^{\sigma}
(\partial_{\sigma} U^{-1}(x) ) U(x) \right\rbrack g^{-1}(x),
\ee
\be \label{3.41}
F_{\mu\nu}^{\sigma} \to F_{\mu\nu}^{\prime\sigma}
= \Lambda^{\sigma}_{~\rho} g(x) F_{\mu\nu}^{\rho} g^{-1}(x),
\ee
\be \label{3.42}
G^{\alpha}_{\mu} \to G^{\prime\alpha}_{\mu}
= \Lambda^{\alpha}_{~\beta} g(x) G^{\beta}_{\mu} g^{-1}(x),
\ee
\be \label{3.43}
G_{\alpha}^{-1\mu} \to G_{\alpha}^{\prime -1 \mu}
= \Lambda_{\alpha}^{~\beta} g(x) G_{\beta}^{-1\mu} g^{-1}(x),
\ee
\be \label{3.431}
g_{\alpha\beta} \to g'_{\alpha\beta}
= \Lambda_{\alpha}^{~\alpha_1} \Lambda_{\beta}^{~\beta_1}
\cdot g(x)  g_{\alpha_1 \beta_1} g^{-1}(x),
\ee
\be \label{3.44}
{\mathbf A}_{\mu\nu} \to {\mathbf A}'_{\mu\nu}
= g(x) {\mathbf A}_{\mu\nu} g^{-1}(x),
\ee
\be \label{3.45}
J(C) \to J'(C') = J \cdot g(x) J(C) g^{-1}(x).
\ee
Action $S$ is invariant local $GSU(N)$ gauge transformation.  \\

\section{Interactions}
\setcounter{equation}{0}

The lagrangian density ${\cal L}$ can also be separated into two 
parts: the free lagrangian density ${\cal L}_F$ and interaction 
lagrangian density ${\cal L}_I$
\be \label{3.46}
{\cal L} = {\cal L}_F + {\cal L}_I,
\ee
where
\be \label{3.47}
\ba{rcl}
{\cal L}_F &=& - \bar{\psi} ( \gamma^{\mu}
\partial_{\mu}  + m ) \psi
- \frac{1}{4} \eta^{\mu \rho} \eta^{\nu \sigma}
\eta_{\alpha \beta } F_{0 \mu \nu}^{\alpha} F_{0 \rho
\sigma}^{\beta}  \\
&&\\
&& - \frac{1}{4} \eta^{\mu \rho} \eta^{\nu \sigma}
A^a_{0 \mu \nu} A^a_{0 \rho \sigma}  ,
\ea
\ee
\be \label{3.48}
\ba{rcl}
{\cal L}_I &=& - ( J(C) - 1 )
 \bar{\psi} ( \gamma^{\mu} \partial_{\mu}  + m ) \psi  \\
 &&\\
&& - \frac{1}{4} \eta^{\mu \rho} \eta^{\nu \sigma}
(J(C) g_{\alpha\beta} - \eta_{\alpha \beta } )
F_{0 \mu \nu}^{\alpha} F_{0 \rho
\sigma}^{\beta}  \\
&&\\
&& - \frac{1}{4} \eta^{\mu \rho} \eta^{\nu \sigma}
(J(C) - 1 )
A^a_{0 \mu \nu} A^a_{0 \rho \sigma} \\
&&\\
&& + g J(C) \bar\psi \gamma^{\mu} (\partial_{\alpha} \psi)
C_{\mu}^{\alpha}
+ i g_s J(C) \bar\psi \gamma^{\mu} T_a \psi
A^a_{\mu}  \\
&&\\
&& + g  \eta^{\mu\rho} \eta^{\nu\sigma} J(C)
(\partial_{\mu} A^a_{\nu} -  \partial_{\nu} A^a_{\mu})
C_{\rho}^{\alpha} (\partial_{\alpha} A^a_{\sigma} )\\
&&\\
&& - \frac{g_s}{2}  \eta^{\mu\rho} \eta^{\nu\sigma} J(C)
f_{abc} A^b_{\rho} A^c_{\sigma}
(\partial_{\mu} A^a_{\nu} -  \partial_{\nu} A^a_{\mu}) \\
&&\\
&& + g \eta^{\mu\rho} \eta^{\nu\sigma} g_{\alpha\beta} J(C)
(\partial_{\mu} C_{\nu}^{\alpha} - \partial_{\nu} C_{\mu}^{\alpha} )
C_{\rho}^{\gamma} (\partial_{\gamma} C_{\sigma}^{\beta} )\\
&&\\
&& - \frac{g}{2} \eta^{\mu\rho} \eta^{\nu\sigma} J(C)
G^{-1 \lambda}_{\alpha} A_{\lambda}^a
A^a_{\mu\nu} F_{\rho\sigma}^{\alpha} \\
&&\\
&& -\frac{g_s^2}{4} \eta^{\mu\rho} \eta^{\nu\sigma} J(C)
f_{abc} f_{ab_1c_1} A^b_{\mu} A^c_{\nu}
A^{b_1}_{\rho} A_{\sigma}^{c_1} \\
&&\\
&&+ g g_s \eta^{\mu\rho} \eta^{\nu\sigma} J(C)
f_{abc} A^b_{\mu} A^c_{\nu} C_{\rho}^{\alpha}
(\partial_{\alpha} A^a_{\sigma} ) \\
&&\\
&& -\frac{g^2}{2} \eta^{\mu\rho} \eta^{\nu\sigma} J(C)
(C_{\mu}^{\alpha} \partial_{\alpha} A^a_{\nu}
-  C_{\nu}^{\alpha} \partial_{\alpha} A^a_{\mu})
C_{\rho}^{\beta} (\partial_{\beta} A^a_{\sigma} ) \\
&&\\
&& -\frac{g^2}{2} \eta^{\mu\rho} \eta^{\nu\sigma}
g_{\alpha\beta} J(C)
(C_{\mu}^{\gamma} \partial_{\gamma} C^{\alpha}_{\nu}
-  C_{\nu}^{\gamma} \partial_{\gamma} C^{\alpha}_{\mu})
C_{\rho}^{\delta} (\partial_{\delta} C^{\beta}_{\sigma} ) \\
&&\\
&& - \frac{g^2}{4} \eta^{\mu\rho} \eta^{\nu\sigma} J(C)
G^{-1 \lambda}_{\alpha} G^{-1 \kappa}_{\beta}
A^a_{\lambda} A^a_{\kappa}
F_{\mu\nu}^{\alpha} F_{\rho\sigma}^{\beta}.
\ea
\ee
In above relations, $A^a_{0\mu\nu}$ and $F_{0\mu\nu}^{\alpha} $ 
are defined by
\be \label{3.49}
A^a_{0\mu\nu} = (\partial_{\mu} A^a_{\nu})
- (\partial_{\nu} A^a_{\mu}).
\ee
\be \label{3.4901}
F_{0\mu\nu}^{\alpha} = (\partial_{\mu} C_{\nu}^{\alpha})
-(\partial_{\nu} C_{\mu}^{\alpha}).
\ee
The explicite expression for $J(C)$ is
\be
J(C) = 1 + \sum_{m=1}^{\infty} \frac{1}{m!}
\left( \sum_{n=1}^{\infty}
\frac{g^n}{n}{\rm tr} (C^n)
\right)^m
\label{3.50}
\ee
From eq.(\ref{3.47}), we can write out propagators of Dirac field,
$SU(N)$ non-Abelian gauge field and gravitational gauge field. From
eq.(\ref{3.48}), we can write out Feynman rules for various interaction
vertexes and calculate Feynman diagrams for various interaction
precesses. We can also see that, because of the influence of the
factor $J(C)$,  matter fields can directly
couple to arbitrary number of gravitational gauge field, which
is important for the renormalization of the theory. 
 \\

\section{Equations of Motion and Energy-Momentum Tensor}
\setcounter{equation}{0}

The equation of motion of Dirac field is
\be \label{3.50}
(\gamma^{\mu} {\mathbf D}_{\mu} + m  ) \psi =0.
\ee
The equation of motion of $SU(N)$ gauge field is
\be \label{3.51}
\partial^{\mu} {\mathbf A}^a_{\mu\nu}
= - g_s \eta_{\nu\sigma} J^{\sigma}_{a},
\ee
where
\be \label{3.52}
\ba{rcl}
J_a^{\nu} &=& i \bar\psi \gamma^{\nu} T_a \psi
+ \eta^{\nu\rho} \eta^{\nu_1 \sigma} f_{abc}
A^c_{\nu_1} {\mathbf A}^b_{\rho\sigma}
- \frac{g}{g_s} \eta^{\mu_1 \rho} \eta^{\nu\sigma}
\partial_{\mu} (C_{\mu_1}^{\mu} {\mathbf A}^a_{\rho\sigma} )\\
&&\\
&& - \frac{g}{2 g_s} \eta^{\mu_1 \rho} \eta^{\nu_1 \sigma}
G^{-1 \nu}_{\sigma_1} F^{\sigma_1}_{\mu_1 \nu_1}
{\mathbf A}^a_{\rho\sigma}
 + \frac{g}{g_s} \eta^{\mu\rho} \eta^{\nu\sigma}
G^{-1 \lambda}_{\tau} (D_{\mu} C_{\lambda}^{\tau} ) 
{\mathbf A}_{\rho\sigma}^a.
\ea
\ee
$J_a^{\nu}$ is a conserved current,
\be \label{3.53}
\partial_{\nu} J_a^{\nu} = 0.
\ee
When gravitational gauge field vanishes, the above current
$J_a^{\nu}$ returns to the conventional current in traditional
non-Abelian $SU(N)$ gauge field theory, which is
\be \label{3.54}
J_a^{\nu} = i \bar\psi \gamma^{\nu} T_a \psi
+ \eta^{\nu\rho} \eta^{\mu \sigma} f_{abc}
A^c_{\mu}  A^b_{\rho\sigma}.
\ee
But if gravitational gauge field does not vanish, 
because of the influence from gravitational gauge field, 
the conventional current eq.(\ref{3.54}) is no longer a 
conserved current.
\\

The equation of motion of gravitational gauge field is
\be \label{3.55}
\partial^{\mu} (\eta^{\nu \sigma} g_{\alpha\beta}
F_{\mu\sigma}^{\beta} )
= -g T_{g\alpha}^{\nu},
\ee
where $T_{g\beta}^{\sigma}$ is the gravitational energy-momentum
tensor
\be \label{3.56}
\ba{rcl}
T_{g\alpha}^{\nu}&=& \bar\psi \gamma^{\nu} \partial_{\alpha} \psi
- \eta^{\mu_1 \rho} \eta^{\nu \sigma} g_{\alpha_1 \beta}
F^{\beta}_{\rho\sigma} (\partial_{\alpha} C_{\mu_1}^{\alpha_1})
- \eta^{\mu_1\rho} \eta^{\nu\sigma} {\mathbf A}^a_{\rho\sigma}
(\partial_{\alpha} A^a_{\mu_1})  \\
&&\\
&& + G^{-1 \nu}_{\alpha} {\cal L}_0 
- \frac{1}{2} \eta^{\mu\rho} \eta^{\lambda\sigma}
g_{\alpha\beta} G^{-1 \nu}_{\gamma}
F^{\beta}_{\mu\lambda} F^{\gamma}_{\rho\sigma}\\
&&\\
&&-\eta^{\mu_1 \rho} \eta^{\nu \sigma}  
\partial_{\mu}( g_{\alpha \beta}  
{C_{\mu_1}^{\mu} F^{\beta}_{\rho\sigma}})
+ \eta^{\mu_1 \rho} \eta^{\nu \sigma}
\partial_{\mu} ( G^{-1 \lambda}_{\alpha} G^{\mu}_{\mu_1}
A^a_{\lambda} {\mathbf A}^a_{\rho\sigma} ) \\
&&\\
&& +\eta^{\mu \rho} \eta^{\nu \sigma} g_{ \alpha \beta}
G^{-1 \lambda}_{\tau} (D_{\mu} C^{\tau}_{\lambda} )
F^{\beta}_{\rho\sigma}
+ g \eta^{\mu\rho} \eta^{\nu \sigma}
G^{-1 \lambda_1}_{\tau} (D_{\mu} C^{\tau}_{\lambda_1} ) 
G^{-1 \lambda}_{\alpha} A^a_{\lambda} {\mathbf A}^a_{\rho\sigma}\\
&&\\
&& - \frac{g}{2} \eta^{\mu_1 \rho} \eta^{\nu_1 \sigma}
G^{-1 \nu}_{\sigma_1} G^{-1 \lambda}_{\alpha}
A^a_{\lambda} {\mathbf A}^a_{\rho\sigma}
F^{\sigma_1}_{\mu_1 \nu_1}
- g \eta^{\mu_1 \rho} \eta^{\nu \sigma}
G^{-1 \lambda }_{\sigma_1} A^a_{\lambda}
(\partial_{\alpha} C_{\mu_1}^{\sigma_1} )
{\mathbf A}^a_{\rho\sigma}.
\ea
\ee
The global gravitational gauge symmetry of the system gives out
another energy-momentum tensor which is called inertial
energy-momentum tensor,
\be \label{3.57}
\ba{rcl}
T^{\mu}_{i \alpha} &=& J(C) \left(
\bar\psi \gamma^{\nu} G^{\mu}_{\nu} (\partial_{\alpha} \psi)
+ \eta^{\mu_1 \rho} \eta^{\nu \sigma}
G^{\mu}_{\mu_1} {\mathbf A}^a_{\rho\sigma}
(\partial_{\alpha} A^a_{\nu})  \right.
 + \eta^{\mu \rho} \eta^{\nu \sigma}  g_{ \beta\gamma}
F^{\gamma }_{\rho\sigma} (\partial_{\alpha} C_{\nu}^{\beta})
+ \delta^{\mu}_{\alpha} {\cal L}_0 \\
&&\\
&&\left. - g \eta^{\mu_1 \rho} \eta^{\nu \sigma} g_{\beta\gamma}
C_{\mu_1}^{\mu} F^{\gamma }_{\rho\sigma}
(\partial_{\alpha} C_{\nu}^{\beta})
+ g\eta^{\mu_1 \rho} \eta^{\nu \sigma} G^{-1 \lambda}_{\beta}
G^{\mu}_{\mu_1} A^a_{\lambda} {\mathbf A}^a_{\rho\sigma}
(\partial_{\alpha} C^{\beta}_{\nu}) \right).
\ea
\ee
Compare eq.(\ref{3.56}) with eq.(\ref{3.57}), we can see that the
inertial energy-momentum tensor is not equivalent to the
gravitational energy-momentum tensor. In this case, they
are not equivalent even when gravitational field vanishes.
When gravitational field vanishes, the gravitational 
energy-momentum tensor becomes $T^{\nu}_{0 g \alpha}$,
\be \label{3.58}
T_{0g\alpha}^{\nu} = \bar\psi \gamma^{\nu} \partial_{\alpha} \psi
- \eta^{\mu\rho}\eta^{\nu\sigma} A^a_{\rho\sigma}
(\partial_{\alpha} A^a_{\mu})
+ \eta^{\nu}_{1\alpha} {\cal L}_0
+ \eta^{\nu\sigma} \partial^{\mu}
(A^a_{\alpha} A^a_{\mu\sigma} ),
\ee
while inertial energy-momentum tensor becomes $T^{\nu}_{0 i \alpha}$
\be \label{3.59}
T_{0i\alpha}^{\nu} = \bar\psi \gamma^{\nu} \partial_{\alpha} \psi
- \eta^{\mu\rho}\eta^{\nu\sigma} A^a_{\rho\sigma}
(\partial_{\alpha} A^a_{\mu})
+ \delta^{\nu}_{\alpha} {\cal L}_0.
\ee
Therefore, we have
\be \label{3.60}
T_{0g\alpha}^{\nu} = T^{\nu}_{0i\alpha}
+ \eta^{\nu\sigma} \partial^{\mu}
(A^a_{\alpha} A^a_{\mu\sigma} ).
\ee
But this difference has no contribution on energy-momentum. The spatial
integration of time component of energy-momentum tensor gives out
energy-momentum of the system. The inertial energy-momentum
$P_{0 i \alpha}$ is
\be \label{3.61}
P_{0 i \alpha} = \int {\rm d}^3 x T^{0}_{0i\alpha},
\ee
and the gravitational energy-momentum $P_{0 g \alpha}$ is
\be \label{3.62}
P_{0 g \alpha} = \int {\rm d}^3 x T^{0}_{0g\alpha}.
\ee
So, their difference is
\be \label{3.63}
P_{0 g \alpha} - P_{0 i \alpha} = - \int {\rm d}^3 x
\partial_i (A^a_{\alpha} A^a_{i 0} )=0.
\ee
It means that, when gravitational gauge field vanishes, equivalence
principle holds. \\

\section{Summary}

In this paper, we have studied unifications of ordinary $SU(N)$
gauge interactions with gravitational gauge interactions, 
which is unified
in the semi-direct product group $GSU(N)$. Because generators
of ordinary $SU(N)$ group and generators of gravitational gauge
group have different dimensions, that is, generators of $SU(N)$
group are dimensionless while generators of gravitational gauge
group have length dimension, it is hard to unify $SU(N)$ gauge
interactions and gravitational gauge interactions in a simple
group. Because of the difference of dimensions of generators,
we need at least two independent parameters for
coupling constant in any kind of unified theory. Because when
we unify $SU(N)$ gauge interactions and gravitational gauge
interactions in $GSU(N)$ group, we only need two independent
parameters for coupling constant, this unified theory can
be regarded as a minimal theory of unification. 
It is impossible to unify four kinds of fundamental 
interactions in a simple group in which
only one independent coupling constant is used. \\

Because $SU(N)$ gauge group and gravitational gauge group
are unified in a semi-direct product 
group, not in a direct product group,
field strength of gravitational gauge field joins into
the definition of gauge covariant field strength of
$SU(N)$ gauge field. This will cause additional interactions
between $SU(N)$ gauge fields and gravitational gauge field,
which even cause that gravitational energy-momentum tensor
is not equivalent to inertial energy-momentum tensor when
gravitational field vanishes, but this difference does not
affect the equivalence of gravitational mass and inertial
mass when gravitational field vanishes. \\

\end{document}